\documentclass[runningheads]{svmult}

\usepackage{makeidx}   
\usepackage{graphicx}  
\usepackage{subeqnar}  
\usepackage{multicol}  
\usepackage{physprbb}  
\makeindex             

\newcommand{\greeksym}[1]{{\usefont{U}{psy}{m}{n}#1}}
\newcommand{\umu}{\mbox{\greeksym{m}}}

\def\etal{{\sl et~al.}}
\def\spose#1{\hbox to 0pt{#1\hss}}
\def\lta{\mathrel{\spose{\lower 3pt\hbox{$\mathchar"218$}}
     \raise 2.0pt\hbox{$\mathchar"13C$}}}
\def\gta{\mathrel{\spose{\lower 3pt\hbox{$\mathchar"218$}}
     \raise 2.0pt\hbox{$\mathchar"13E$}}}

\def\kms{{\rm\,km\,s^{-1}}}

\def\kpc{{\rm\,kpc}}

\def\msun{{\rm\,M_\odot}}

\begin{document}
\title*{Modelling Kinematics and Dark Matter: \protect\newline
The Halos of Elliptical Galaxies\footnote{To appear in: Planetary
Nebulae beyond the Milky Way, ESO Astrophysics Symposia, eds.\
Stanghellini, L., Walsh, J. R., Douglas, N. G., Springer, 
Heidelberg.}}
\toctitle{Modelling Kinematics and Dark Matter:
\protect\newline The Halos of Elliptical Galaxies}
%
%
\titlerunning{Modelling the Halos of Elliptical Galaxies}
%
\author{Ortwin Gerhard\inst{ }}
%
\authorrunning{Ortwin Gerhard}
%
%
\institute{Astronomisches Institut,
     Universit\"at Basel,
     Venusstrasse 7,
     CH-4102 Switzerland
}

\maketitle              

\begin{abstract}
This review is focussed on the outer halos of elliptical galaxies.
Its emphasis is on (i) planetary nebulae as test particles to trace the 
stellar kinematics at large radii, (ii) the observed angular momentum
in elliptical galaxy halos and its theoretical relevance, (iii) dynamical
modelling of stellar-kinematic data, and (iv) a discussion of the
evidence for dark matter halos in ellipticals from a wide range of
measurements.
\end{abstract}

\section{Introduction}
In current hierarchical theories, elliptical galaxies represent an
advanced stage of the galaxy formation process: they are massive
galaxies found in dense environments. They reflect `early' conditions
in galaxy formation as well as subsequent merging processes,
making the study of their properties particularly interesting.

However, their old stellar populations and complicated orbital
structure make many important questions harder to answer for
ellipticals than for spirals. This includes cosmological questions
about their ages and precise formation mechanism, about the
concentration of their dark matter halos, and the segregation of
baryonic matter, as well as dynamical questions about the orbit
distribution and total angular momentum, to name only a few.

The subject of this review is the halos of elliptical galaxies and how
planetary nebulae observations and dynamical modelling can help answer
some of these questions.  Several other papers in this session
elaborate on the topics discussed here.

\section{Planetary Nebulae as Kinematic Tracers} 

Planetary Nebulae (PNe) are excellent test particles to trace the
kinematics in the outer halos of elliptical galaxies. They occur as a
brief stage in the late evolution of stars with masses between
$\sim0.8-8\msun$, when these stars evolve from the asymptotic giant
phase to their final white dwarf stage. Their nebular gas envelope
converts up to 15\% of the central star's radiation energy to photons
in the characteristic $\lambda$5007 [OIII] emission line [1].  From
this emission line, PNe can be identified and their radial velocities
measured spectroscopically at distances of up to 20--30 Mpc. With 8m
class telescopes [2] or the special-purpose Planetary Nebula
Spectrograph (PN.S) [3], several hundred PNe have been found in giant
elliptical galaxies with distance up to the Virgo cluster. The PN
number density approximately follows the surface density of stars;
thus a few tens of PNe can be used to measure the stellar kinematics
at several effective radii $R_e$, where the surface brightness is too
faint for absorption line spectroscopy (ALS).

From these PN velocities, constraints on the angular momentum,
dynamics, and mass distribution in the outer halos of ellipticals can
be derived; these are important to compare with predictions from
theories of galaxy formation. 

In this analysis, it is always advantageous if additional information
can be included. Especially, stellar kinematics from ALS are highly
desirable (i) to determine the mass-to-light ratio in the inner,
presumably baryon-dominated parts of the galaxy, and (ii) to narrow
down the range of permitted dynamical anisotropies at 1-2 effective
radii [4]. Also, globular cluster (GC) velocities are now becoming
available in substantial numbers around luminous ellipticals. While
the spatial distribution and dynamics of the globular cluster system
may be different from those of the stars, analysis of these velocities
gives additional constraints on the common gravitational potential [5].
Lastly, an ideal case is when the galaxy potential is narrowly
constrained from Chandra or XMM X-ray observations [6].  Then the
information obtained with the PN kinematics can be used entirely for
inferring the orbital structure of the outer stellar halo in the known
potential.  These issues are further discussed later; the present section
is concerned with PNe as kinematic tracers.

\subsection{Does the PN Number Density Follow Light?}

Although PNe are representative for the bulk of all stars,
there are some population effects that can bias their number
density profile with respect to the galaxy's luminosity profile. The
PN luminosity function (PNLF) is observed to have a near-universal
form and bright cut-off magnitude [7]. Then the quantity
$\alpha_{X,n}$ is defined as the ratio of the number of PN in a
stellar population, $n$ magnitudes down the PNLF from the cut-off
magnitude, to its total stellar luminosity, in a specified
wavelength band $X$.  While the bolometric $\alpha_{\rm bol,\infty}$
is expected to be nearly independent of stellar population [8], Hui
\etal\ show that the blue $\alpha_{B,2.5}$ depends on colour B-V,
decreasing by a factor of 4 from the M31 bulge to the reddest
elliptical galaxies [9]. Age, or metallicity, or both, must influence
$\alpha_{B,2.5}$. Theoretical models predict that the brightest PNe in
B have central stars with progenitor masses $\sim 2.5\msun$ [10], with
luminosity also dependent on uncertain post-AGB evolution parameters
and metallicity. The presence of a younger population in elliptical
galaxies, whether made in situ or accreted, could thus change the
number of bright PNe significantly. Real ellipticals can have colour
and metallicity gradients.  Therefore it is prudent to check carefully
whether the observed (usually magnitude-limited) PNe samples do indeed
trace light.

Observational tests of this question have been carried out in a number
of galaxies, including M31, Cen A, the Leo group galaxies NGC 3377,
3379 and 3384, several Virgo ellipticals, the Fornax galaxies NGC 1399
and NGC 1404, and NGC 4697 [2,7,9,11]. Generally, the number of PNe as
a function of isophote follows the luminosity distribution for radii
outside $\sim 1-2'$, in some cases very well. At smaller radii some
PNe are lost against the bright background of the galaxy, and the PNe
no longer track the luminosity profile. A colour gradient such that
the inner parts of the galaxy are redder with fewer PN per blue light
may further decrease the number of PNe per unit light there; NGC 4697,
for example, has a slight gradient of this kind [12].

\subsection{Do the PN Kinematics Follow the Absorption Line
Stellar Kinematics?}

If the distribution of PNe follows light, the PNe can be considered as
faithful tracers also for the stellar kinematics. A test of this is to
compare the rotation velocities $v$ and velocity dispersions $\sigma$
obtained from PN radial velocities with those from ALS, taking into
account the greater smoothing of the PN velocity field due to spatial
averaging. Mendez used his sample of 535 PN velocities in NGC 4697 for
a simple comparison [2].  His sample, restricted to a stripe around the
major axis, indeed followed the rotation seen in the ALS data;
however, the number of PNe per mean velocity point is $\sim 10$ and so
the errors are large, $\sim 50\kms$. Similarly, the PN velocity
dispersions in major axis sectors were consistent with the ALS
dispersions within $\sim 20\kms$.

In such a case the PN velocities and the ALS velocity measurements can
be simply combined in modelling the stellar kinematics. Even the
incompleteness in the center does not present a problem, because the
probability of losing a PN against the bright galaxy background is
uncorrelated with its radial velocity. Then only the PN velocities,
but not their radial distribution must be used in the modelling.  If,
on the other hand, because of an outward colour gradient and the
observed dependence of $\alpha_B$ on colour, the distribution of PNe
is more extended than the stellar light, say, then the PNe velocity
dispersions would overestimate the stellar velocity dispersions, as
can be seen most easily from the spherical Jeans equation.  In this
case, the PNe would have to be included in the modelling as a separate
test particle distribution.

Another kinematic bias can occur for elliptical galaxies in galaxy
clusters, in that the PN samples, particularly in the outer galactic
halo, can be contaminated by intracluster PNe [13]. ICPNe would be
difficult to disentangle from galactic PNe when the velocity
dispersions of the galaxy and its host cluster are comparable, such as
for NGC 1399 in the Fornax cluster.

\subsection{How Many PNe Velocities Are Needed for Orbital
and Potential Analysis?}

With a few tens of PN velocities it is possible to detect a clear
trend of rotation in the outer halo; see, e.g., the work of Arnaboldi
\etal\ for NGC 4406 and NGC 1316 [13, 14]. To determine the detailed shape of
the rotation velocity field requires substantially larger samples.
Fig.~1, due to N.~Sambhus, shows the results of fitting the velocities
of approximately 200 PNe from PN.S observations [15] in the E0/1
galaxy NGC 3379 with a non-parametric spline method. The velocity
field shows a peak-to-peak rotation of $\sim 100\kms$ at $\sim 100''$,
with a subsequent decline out to $\sim 300''$, confirmed by the
rotation along the line of maximum gradient. However, both the
detailed shape of the rotation curve and the reality of the apparent
asymmetry in the velocity field remain uncertain. For comparison,
SAURON data show a clear rotation gradient in the central 20'', which
is unresolved in the PN velocities [16]. The conclusion from this is
that PN velocities should be used as an outward extension of the ALS
velocity measurements whenever possible [14].

\begin{figure}[b]
\begin{center}
\includegraphics[width=.88\textwidth]{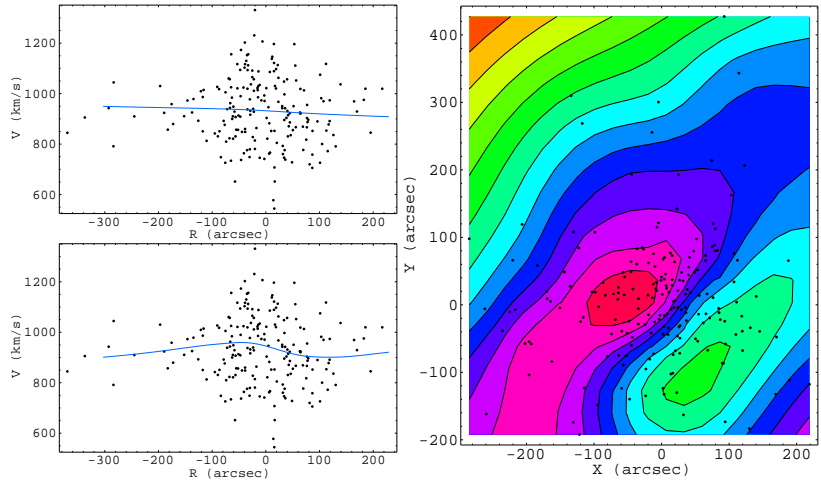}
\end{center}
\caption[]{Velocity field and streaming velocities from $\sim 200$ PNe 
in NGC 3379.  The right-hand figure shows a smoothed spline model
for the projected mean velocity, derived from the PN velocities; 
contour spacing is $10\kms$ and the
galaxy center is at (0,0). The two plots on the left show fits with
different smoothing to the distribution of PN velocities versus
distance along the direction of maximum gradient.}
\label{eps1}
\end{figure}

The second important application of PN velocities is to help determine
the masses of ellipticals in their outer regions, where the dark
matter halos should begin to dominate.  Because of the large freedom
in the orbital structure and velocity anisotropy in elliptical
galaxies, this is not an easy task. If no other kinematic constraints
are available, Merritt has estimated that at least many hundreds of PN
velocities would be required [17]. Thus, it is again advantageous to use the
PN data in conjunction with ALS velocities and other data. From ALS
data out to $\sim 2R_e$, including line profile shape parameters $h_3$
and $h_4$, a range of dynamical models and gravitational potentials
can be determined that are consistent with these data. Within these
models, the implied orbital anisotropy and halo mass are correlated.
Then only a few tens of PN velocities beyond $2R_e$ are needed to
further narrow down the permitted dynamics and mass distribution [4].

\section{Angular Momentum in the Outer Parts of Ellipticals}

One of the most interesting early results obtained with PNe was the
discovery of rapid rotation in the halos of the giant elliptical
galaxies Centaurus A, NGC 1399, and NGC 1316 [18, 14]. The PNe radial
velocity fields in these early-type galaxies show that their outer
halos are rapidly rotating, and that these galaxies contain comparable
angular momentum to spiral galaxies of similar luminosity. Recent work
on several further ellipticals has shown that most of these do not
rotate rapidly at large radii; see diagram in [19]. There appears to
be a range of outer rotation properties observed.

The amount of rotation in the halos of elliptical galaxies is a
valuable probe for how these systems formed. Elliptical galaxies are
now believed in general terms to form by merging processes; what is
less clear is the kind of progenitors that dominated in the formation
of the present population of ellipticals. Spiral-spiral galaxy
mergers, observed in the local universe and spurring Toomre's original
merger hypothesis [20], give remnants that morphologically and
kinematically resemble elliptical galaxies in many ways
[21]. Equal-mass mergers have low $v/\sigma\sim0-0.2$ within $R\lta
R_e$, as observed for giant ellipticals, due to angular momentum
transfer from inner regions to the extended outer halos by dynamical
friction in the merger. These remnants can, however, contain
significant angular momentum at large radii, reaching
$v/\sigma\sim0.2-0.5$, even though a lot of spin angular momentum is
carried away by material in the tidal tails. Unequal-mass mergers
rotate faster than equal-mass mergers.  As argued in [21], binary
mergers of disk galaxies may be the main formation mechanism of low-
and intermediate mass ellipticals.

An alternative merging channel to form an elliptical is through
multiple major and minor mergers in a compact group of galaxies
[22]. In this case the tidal forces are more effective in disrupting
the progenitors before coalescence, so dynamical friction is less
effective. As a result, the remnants have more angular momentum in
their inner parts than spiral-spiral mergers, placing them not far
from the oblate-isotropic line in the $v/\sigma-\epsilon$ diagram, and
their outer parts may reach $v/\sigma\sim 1$.

Computations of the angular momentum of dark matter halos growing by
merging and accretion in hierarchical universes result in low spin 
parameter
$\lambda$ and low $v/\sigma$ [23]. Interestingly, the values of
these parameters in the evolution of individual halos are most likely
to increase in major mergers, and generally decrease in multiple
accretion of satellites.  If this is indicative for the luminous
components also, then ellipticals that were last shaped by a major
merger should contain the highest angular momenta. The rotation
velocities in the remnant halos are fairly constant with radius,
however, confirming that the dissipation and dynamical friction
processes acting specifically on the baryonic component are crucial
for shaping the angular momentum distributions in elliptical galaxies.

These results show that there is no simple, one-to-one correspondence
between angular momentum at large radii and formation mechanism.  For
example, of the ellipticals with outer PN kinematics, Centaurus A is
believed to have formed from the merger of two disk galaxies; in this
galaxy $v/\sigma$ rises to $\sim 1$ beyond $R=15\kpc$ [18]. However,
how much angular momentum resides in the outer halos of elliptical
galaxies is clearly a key issue which, when understood for a
representative sample of elliptical galaxies, will be crucial for
determining the merging channel that dominated their formation. This
is a research program that can ideally be tackled by PNe radial velocity
measurements with the PN.S.

\section{Dynamical Analysis of Kinematic Data}

This Section gives a brief overview of how the gravitational potential
of an elliptical galaxy can be inferred from stellar-kinematic
data. These data may include PN velocities, which individually sample
the stellar line-of-sight velocity distributions $L(v)$ (hereafter
LOSVD) at their positions in the galaxy image, as well as ALS
spectroscopy measurements of the first few moments of the LOSVD, at a
set of 1D or 2D binned positions. 

As is well known, velocity dispersion profile measurements (and
streaming velocities, if the galaxy rotates) do not suffice to
determine the distribution of mass with radius, due to the degeneracy
with orbital anisotropy. With very extended measurements a constant
$M/L$ model can be ruled out (e.g., [18]), but the detailed $M(r)$
still remains undetermined. Thus mass determination in elliptical
galaxies always involves determining the orbital structure at the same
time, and requires a lot of data. LOSVDs from absorption line profile
shapes provide additional data with which the degeneracy between orbit
structure and mass can largely be broken.  Simple spherical models are
useful to illustrate this [24]: at large radii, radial orbits are seen
side-on, resulting in a peaked LOSVD (positive Gauss-Hermite parameter
$h_4$), while tangential orbits lead to a flat-topped or double-humped
LOSVD ($h_4<0$). Similar considerations can be made for edge-on or
face-on disks [25] and spheroidal systems [26].

One may think of the LOSVDs constraining the anisotropy, after which
the Jeans equations can be used to determine the mass distribution.
However, the gravitational potential influences not only the widths,
but also the shapes of the LOSVDs (see illustrations in [24]).
Furthermore, eccentric orbits visit a range of galactic radii and may
therefore broaden a LOSVD near their pericentres as well as leading to
outer peaked profiles. Thus, in practice, the dynamical modelling to
determine the orbital anisotropy and gravitational potential must be
done globally, and is typically done in the following steps:

(0) choose geometry (spherical, axisymmetric, triaxial);

(1) choose dark halo model parameters, and set total luminous plus
    dark matter potential $\Phi$;

(2) write down a composite distribution function (DF)
    $f=\Sigma_k a_k f_k$, where the $f_k$ can be orbits, or
    DF components such as $f_k(E,L^2)$, with free $a_k$; 

(3) project the $f_k$ to observed space, $p_{jk}=\int K_j f_k d\tau$,
    where $K_j$ is the projection operator for observable $P_j$, and
    $\tau$ denotes the line-of-sight coordinate and the velocities; 
    
(4) fit the data $P_j = \Sigma_k a_k p_{jk}$ for all observables
    $P_j$ simultaneously, minimizing a $\chi^2$ or negative
    likelihood, and including regularization to avoid spurious large
    fluctuations in the solution. This determines the $a_k$, i.e, the
    best DF $f$, given $\Phi$, which must be $f>0$ everywhere;

(5) vary $\Phi$, go back to (1), and determine confidence limits
    on the parameters of $\Phi$. \\
If the mass distribution and gravitational potential are known from
analysis of, e.g.,  X-ray data, step (5) can be omitted. Because this eliminates
the degeneracy with orbital anisotropy, considerably fewer data are
then needed. 

Such a scheme has been employed regularly to model spherical and
axisymmetric galaxies [e.g., 4,27], using mostly orbits
(``Schwarzschild's method'') and, more rarely, distribution function
components. Discrete velocity data were modelled in [e.g., 5,17].
The modelling techniques used to constrain black hole masses from
nuclear kinematics and dark halo parameters from extended kinematics
are very similar.

Line-profile shape parameter measurements are now available for many
nearby ellipticals, but those reaching to $\sim 2 R_e$ are still
scarce [e.g., 28, 4].  Modelling of the outer mass profiles of
ellipticals from such data has been done for some two dozen round
galaxies in the spherical approximation, and for a few cases using
axisymmetric three-integral models.

From recent SAURON integral field measurements it has become clear
that the kinematics of many elliptical galaxies show features
imcompatible with axisymmetry [16]. This has motivated Verolme \etal\
to extend Schwarzschild's method to triaxial models [29], a
development that has only recently become possible with the increase
of available computing power. In parallel, adaptive N-body codes are
being developed that use the algorithm of Syer \& Tremaine for
training N-body systems to adapt to a specified set of data
constraints [30]. These latter models have the additional advantage of
simultaneously providing a check for the model's dynamical
stability. One application has been to the rotating Galactic bar; work
currently in progress is on axisymmetric and rotating triaxial
galaxies.

\section{The Dynamics and Dark Matter Halos of Elliptical Galaxies}

\subsection{Results from Dynamical Analysis of Absorption Line Profiles}

Giant ellipticals with $L\gta L_\ast$ are typically fit best by
dynamical models that have near-isotropic to modestly radially
anisotropic velocity dispersions at intermediate radii, $0.5-1R_e$
[4,27,31].  This is based on spherical and axisymmetric three-integral
models of about two dozen ellipticals. In ellipticals less luminous
than $M_B=-19.5$ rotation is important and, moreover, these galaxies
become rotation-dominated ($v/\sigma\sim 2$) at  $1-2R_e$, similar
to disk-dominated S0 galaxies [32].

Most of the results on dark matter halos in ellipticals obtained from
ALS data to $2R_e$ are based on spherical models for round
ellipticals. [4,31] analyzed the line-profile shapes of a sample of 21
mostly luminous, slowly rotating, and nearly round elliptical galaxies
in a uniform way, using spherical DF models. Intrinsic
deviations from sphericity and embedded, near-face-on disks can play a
role in only a small number of these bright galaxies. Nevertheless, a similar
study using three-integral axisymmetric models of near-edge-on
galaxies avoiding these issues will be very worthwhile. For these
reasons the three-integral models of Matthias \& Gerhard for the
boxy E4 elliptical galaxy NGC 1600 provide some of the strongest
evidence for radial anisotropy, because this galaxy must must be
viewed nearly edge-on [27].

The sample of Kronawitter \etal\ [4] includes a subsample with mostly
new extended kinematic data, reaching to $\sim 2R_e$, and a subsample
based on the less extended older data of [28]. Based on these data and
on photometry, non-parametric spherical models were constructed from
which circular velocity curves, anisotropy profiles, and radial
profiles of $M/L$ were derived, including confidence ranges. 
The circular velocity curve (CVC) for test particles on circular
orbits of varying radius is a convenient measure of the potential,
even though luminous elliptical galaxies do not rotate rapidly. The
CVCs of the elliptical galaxies analysed by [4,31] are flat to within
$\simeq 10\%$ for $R\gta 0.2R_e$ to $R\gta 2R_e$, independent
of luminosity (Fig.~2).  This argues against strong luminosity
segregation in the dark halo potential.

The dynamical models imply small to modest amounts of dark matter
within $2R_e$ [4,31]. However, constant $M/L$ models can be ruled out
only for 7/21 ellipticals in this sample, at the 2$\sigma$ level.
There are ellipticals in this sample which are very well represented
by constant $M/L$ models, and no indication for dark matter within
$2R_e$, and others where the best dynamical models result in local
$M/L_B$s of 15-30 at $2R_e$.  Likewise, Magorrian \& Ballantyne [25]
find evidence for additional dark matter only in a subset of their
galaxies, using constant-anisotropy spherical modelling.  I.e., despite
the uniformly flat CVCs, there is a spread in the ratio of the CVCs
from luminous and dark matter. As in spiral galaxies, the combined
rotation curve of the luminous and dark matter is flatter than those
for the individual components (``conspiracy'').

In the models with maximum stellar mass, the dark matter contributes
$\sim 10-40\%$ of the mass within $R_e$. The flat CVC models, when
extrapolated beyond the range of kinematic data, predict equal
interior mass of dark and luminous matter at $\sim 2-4R_e$, consistent
with results from X-ray analyses.  Even in maximum stellar mass
models, the implied halo core densities and phase-space densities are at least
$\sim 25$ times larger and the halo core radii $\sim 4$ times smaller
than in maximum disk spiral galaxies with the same circular velocity
[33].  This could imply that some elliptical galaxy halos collapsed at high
redshifts or perhaps even that some of their dark matter might be baryonic.

\begin{figure}[t]
\begin{center}
\includegraphics[width=.8\textwidth]{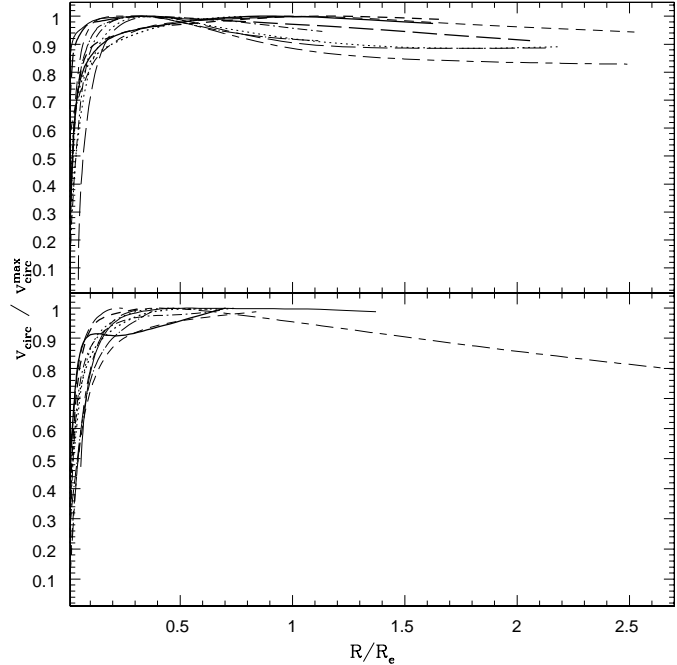}
\end{center}
\caption[]{``Best model'' circular velocity curves of all galaxies
from the sample of [4], plotted as a function of
radius scaled by the effective radius $R_e$, and normalized by the
maximum circular velocity. The upper panel shows the galaxies from the
extended kinematics subsample, the lower panel the galaxies from the
subsample with older data from [28]. (Figure from [31].)}
\label{eps1}
\end{figure}

\subsubsection{Dust and Scattered Light?} 

Baes \& Dejonghe have argued that scattered light from a high-velocity
nucleus might give rise to broad wings in the LOSVD at large radii, in
such a way as to mimic a dark matter halo in dynamical analysis [34].
This requires an optical depth $\tau\simeq 1$, i.e., that elliptical
galaxies are on the verge of opaqueness. No such effect on the surface
density of background galaxies is known to this author. A direct test
is moreover possible when velocity measurements are available
simultaneously from ALS data and from PNe, as the PN velocities are
unaffected (e.g., NGC 4697 above), or when mass determinations are
possible with independent techniques (e.g., NGC 1399 below). Also, the
effect should be absent when the dispersion profile is constant; one
such case in the sample of [4] is NGC 7626.

\subsection{Results from Analyses Including PN or GC Velocities}

Because the dark matter fraction inside $2R_e$ is still modest, and
the orbit structure in the outer main bodies of ellipticals is not
well-constrained by data that end at $2R_e$, it is important to
include discrete velocity data from planetary nebulae (PN) or globular
clusters (GC) that reach to larger radii. 

The results so far are mixed. PNe velocity dispersions in several
intermediate luminosity ellipticals were found to decline with radius,
and to require little if any dark matter at $2-5R_e$ [2,15]. Two of
these (NGC 3379, 4494) were also part of the sample of [4,31], but
were consistent with constant M/L also in that study. NGC 3379 has
been argued to be a face-on S0 and a weakly triaxial system viewed
face-on [35]. This would seem to be not a likely explanation for several
galaxies at the same time, unless selecting round galaxies introduces
a significant bias in this sense. It is comforting that the
near-edge-on, similar luminosity NGC 4697 also has little evidence for
dark matter [2]. See Romanowsky's paper for further discussion
[19]. In Cen A, the observed PN kinematics do require a dark halo,
consistent with the GC velocities, but the implied $M/L_B$ is
noticeably low [18, 36] when compared with values determined from the
hot gas in X-ray bright ellipticals. A similar, mild gradient in
$M/L_B$ was inferred from PN velocities in NGC 1316 [14]. In several
further elliptical galaxies, globular cluster velocities were used to
estimate host galaxy masses, increasing outwards to several effective
radii, but no $M/L_B$ values were given [37].

Two interesting cases are the central galaxies of the Fornax and Virgo
clusters, NGC 1399 and M87. In NGC 1399, the dynamical models for the
ALS kinematics, predicting significantly increased $M/L_B$ already at
$1-2R_e$ [4], the PN and GC velocities [5,18,38], and the ROSAT and
ASCA X-ray data [39] imply a steady outward rise in $M/L_B$. The PN
and GC velocities are in the right radial range to allow a study of
the transition between the potential of the central NGC 1399 galaxy
and the potential of the Fornax cluster.  Similarly, from a study of
the stellar kinematics and the GC velocities around M87, and a
comparison to the X-ray mass profile, Romanowsky \& Kochanek [5, see
also 40] find a rising circular velocity curve, and suggest that the
potential of the Virgo cluster may already dominate at $r \sim 20
\kpc$ from the center of M87. The dark matter distribution inferred by
them is in agreement with that inferred from the X-ray data [6].

\subsection{Results from Gas Rotation Measurements}

Generally, elliptical galaxies do not contain cold gas disks or rings
that can be used to measure their halo circular velocities,  but there
are some notable exceptions. One such example is the early-type 
galaxy IC 2006 in the Fornax cluster. This galaxy has a remarkably
axisymmetric HI ring extending from $0.5-6R_e$, with circular velocity
nearly constant in this range of radii. In this galaxy, $M/L$ rises strongly
with radius, signifying a nearly axisymmetric dark halo [41].

Other ellipticals with settled HI disks are described in [42]. These
galaxies probably acquired the HI gas by accretion. The dynamical
interpretation requires understanding the shape of the potential. In
several of these ellipticals, $M/L_B$ shows a substantial increase
with radius, e.g., NGC 2974, NGC 3108, NGC 4278. In NGC 3108, again
the rotation curve is flat, to $6R_e$, in agreement with the results
from analysing stellar kinematics discussed above.

\subsection{Results from Modelling X-ray Data}

The hot X-ray emitting gas in luminous ellipticals is another means to
estimate the gravitational potential. This requires measuring the
X-ray emissivity and temperature. Typical temperature profiles are
nearly constant. Some problems are that point sources must be removed
from the emission maps, requiring high spatial resolution, and that
the assumed hydrostatic equilibrium may not be strictly
fulfilled. However, any gas motions will have velocities of order
the sound speed, so the
error made with this assumption is not likely to be very large.

X-ray mass measurements have been made regularly as new X-ray missions
were flown, but have tended to concentrate on a small number of
galaxies.  With the new high-spatial and spectral resolution and
high-sensitivity data from Chandra and XMM this situation may soon
change. In the detailed analyses sofar, $M/L$ ratios of $\sim 50$ at
10's of kpc have been found in, e.g., NGC 1399, M87, NGC 4472, NGC
4636, NGC 507, NGC 720 [6,39,43]; these galaxies are often the centers
of galaxy clusters, however.  Ubiquitous halos have furthermore been
found in studies of the family properties of ellipticals from X-ray
data [44].

\subsection{Conclusion}

In addition to the body of evidence reviewed here, studies of
satellite dynamics from SDSS data indicate large $M/L\sim 100-200$ at
radii $260h^{-1}\kpc$ in $0.5-3 L_\ast$ galaxies, and gravitational
lensing studies indicate nearly isothermal mass profiles (constant
CVCs) with a modest dark matter fraction inside $2R_e$ [e.g., 45]. 
Space did not permit a detailed discussion of these issues here.

Alltogether the case is strong that, like all galaxies, elliptical
galaxies have dark matter halos surrounding them, with most inner
circular velocity curves nearly flat. But there may be some
intermediate luminosity objects in which the halos are more diffuse.
Future work to understand this is clearly worthwhile.

%

\end{document}